\documentstyle[prl,aps,psfig]{revtex}
\begin{document}
\draft
\twocolumn[\hsize\textwidth\columnwidth\hsize\csname
@twocolumnfalse\endcsname

\title{Neutral Mutations and Punctuated Equilibrium 
in Evolving Genetic Networks
}

\author{Stefan Bornholdt$^a$ and Kim Sneppen$^b$}

\address{$^a$ Institut f\"ur Theoretische Physik, Universit\"at Kiel, 
Leibnizstrasse 15, D-24098 Kiel, Germany  
\\ 
$^b$ NORDITA, Blegdamsvej 17,  
DK-2100 Copenhagen, Denmark}
\date{Received 25 August 1997}    

\maketitle

\begin{abstract}
Boolean networks may be viewed as idealizations 
of biological genetic networks,
where each node is represented by an on-off switch which is a function of 
the binary output from some other nodes.
We evolve connectivity in a single Boolean network,
and demonstrate how the sole requirement of sequential
matching of attractors may open for an evolution
that exhibits punctuated equilibrium.
\end{abstract} 
\pacs{Published as Phys.\ Rev.\ Lett.\ {\bf 81}, 236 (1998).}  
]

Evolution of life is presumably a random 
process with selection \cite{Darwin}.
It has been discussed whether this process 
can be viewed as some hill climbing process \cite{Wright}, 
or whether evolution mostly
happens as a random walk where changes do not influence the phenotype,
and thus are neutral \cite{Kimura}.
Originally, the case of evolution as adaptation in an externally 
imposed fitness landscape has been proposed 
by Sewall Wright \cite{Wright}, and later formed the basis for models of 
punctuated equilibrium by Newman \cite{Newman} and Lande \cite{Lande}.
The case for neutral evolution has been presented by Kimura \cite{Kimura},
and is experimentally supported on the micro-level by the observation 
that there are many functionally identical variants 
of most of the important macromolecules of life.

The observation of punctuated equilibrium in the fossil record,
recently discussed by Gould and Eldredge \cite{GE1993}, may be taken
as an indication that evolution of a species consists of exaptations
of jumping from one hill top to another nearby in some 
fitness landscape. Naturally such jumps will be rare, 
separated by large time intervals
where species are located at a fitness peak, and the resulting
evolutionary pattern will show punctuations as 
indeed seen in the fossil record. This picture of single species
evolution in a given fixed landscape has been modeled explicitly
by Newman\cite{Newman} and Lande \cite{Lande}. 

However, also neutral evolution may show punctuations 
as, for example, might be visualized by
finding the exit in a labyrinth or from finding a golf hole by 
means of a random walk in a flat landscape.  
The picture here is that genetic changes always take place,
but that the phenotypic changes only rarely occur. 
This has recently been demonstrated by modeling
the evolution of RNA secondary structure by P. Schuster 
and collaborators \cite{Schuster}. 
For these molecules, mutating a single nucleotide often does 
not induce any changes in their secondary structure,
and the mutation is considered neutral. 
Occasionally, however, one mutation can
lead to a complete readjustment of the structure,
usually accompanied by a major change in its functionality.

In any case, as demonstrated by Bak and Sneppen \cite{BS1993},
punctuated equilibrium on the organism level might be connected
to the episodic punctuations observed on the ecosystem level.
The crucial element of such an extrapolation is 
that the environment of each
species depends on species which are ecological neighbors,
thereby allowing punctuations to propagate across the ecosystem.

In the present paper we propose to evolve a single 
genetic network, ideally representing a single species.
The evolution is driven by a noisy environment.
The evolutionary step consists of random mutations 
combined with selection of mutants preserving the phenotype with 
respect to a given environment. Thus, the only requirement 
in this minimalistic model is continuity in phenotype. 
Other changes in genotype are allowed, creating a path of 
neutral mutations. 
We will discuss how this requirement of continuity in 
evolution may constrain and guide the evolution of an individual species 
in the face of a constantly changing environment. 

Our fundamental constituents are the genes of the organism,
and the evolution we consider is on the genetic network level.
Although genetic networks consist of biochemical switches \cite{GSWITCH},
it has been proposed that the on-off nature of these switches
can be well approximated by Boolean functions 
\cite{Bool-Gen,Somogyi,Sales}, 
eventually with asynchronous updating
\cite{Thieffry}.
We here consider networks of random Boolean functions,
idealized by synchronized updating.
The functionality we test for is 
attractors of these networks \cite{Somogyi}.
Boolean networks are known to exhibit a rich dynamical behavior,
including fixed points, periodic attractors and long transients.
Further, the number of attractors, their length, and the
length of the transients strongly depend on the 
connectivity number \cite{Kauffman}.
In this paper we do not address any question about
the timescale of these attractors.
Instead we consider a longer evolutionary timescale
connected to the change in geometry of the networks under mutation. 

We implement continuity in evolution by testing for reaching 
a given attractor on subsequent 
steps, but allowing changes that modify attractors that are not 
tested from the actual initial condition. In subsequent steps, 
the initial condition (modeling the environment) assumes new 
random values which subsequently allow previous neutral 
mutations to surface in the phenotype. 
The philosophy of this subdivision between initial state and 
function is that while the Boolean functions are manifested by 
various DNA binding proteins, the initial state of the system
is set by the chemical composition of the environment.
This environment changes due to conditions beyond the 
control of the Boolean gene regulatory circuit.   

Consider a genetic network with $N$ genes. Each of these genes
can be assigned a Boolean variable $\sigma_i=$ $0$ or $1$.
For each of the $N$ genes we define an updating matrix 
in the form of a lookup table which determines its output 
for each of the possible $2^{N}$ 
input states from the $N$ genes in the system.
This Boolean updating matrix is assigned random values, all rules
are {\em a priori} equally possible. The matrix
is effectively quenched on evolutionary timescales. 
Finally we define which gene is actively connected to which, by a matrix
$w_{ij}$ that defines the input to gene number 
$i$ from gene number $j$ as
$w_{ij}\sigma_j$. The entry value of the connectivity matrix 
$w_{ij}$ can take values $0$, if $i$ is not connected to $j$, 
and $1$, if $i$ is connected to $j$.  
Typically only a fraction of the connectivity matrix entries is in use,
and the average number of connected inputs per gene is called the 
connectivity $K$. It varies between $0$ and $N$, meaning that 
$K$ may include self couplings. Thus $K=0$ means that all is fixed
to the output state specified by input $(0,...,0)$ to all genes.

The system we evolve is the set of couplings $w_{ij}$ in
a single Boolean network. 
The simulation starts with a low but finite connectivity,
here an initial average connectivity of $K=1$ per site.
One evolutionary time step of the network is:
\begin{enumerate} 
\item Select a random input state to the network $\{ \sigma_i \}$.
Iterate the system, called the mother, from this state until a 
final attractor is determined.
\item Create a daughter network by a) adding, b) removing, or c) 
adding and removing a weight in the coupling matrix $w_{ij}$ at random, 
each option occurring with probability $p=1/3$.
Iterate the daughter system from the same initial state 
as that selected for the mother and test whether it reaches 
the same attractor as the mother system did.
In case it does then replace mother with daughter network and go to step 3.
In case another attractor is reached, keep mother network and go to step 3.
\item Then finally one random bit of the total $N*2^N$ lookup table entries  
is flipped to another value. This allows for a convenient
self averaging of the system, and in fact represents a very slow change.
\end{enumerate}   

Iterating these steps makes an evolutionary algorithm that 
represents the sole requirement
of continuity in evolution and how this may proceed under an 
environment that fluctuates. No selective pressure is applied.
Step 3 rarely affects the active part of the network because 
$K$ typically remains low compared to $N$, and thus $2^K<<2^N$.
It is included in order to obtain a self averaging of the system,
that elsewise tends to have small effective statistics
even for the small $N$ we can simulate.

In Fig.\ 1a we show how the connectivity ($K$)
of this system evolves with time in a network of size $N=16$.   
\begin{figure}[htb]
\let\picnaturalsize=N
\def\picsize{85mm}
\def\picfilename{punctuationsfig1a.ps}
\ifx\nopictures Y\else{\ifx\epsfloaded Y\else\input epsf \fi
\let\epsfloaded=Y
\centerline{\ifx\picnaturalsize N\epsfxsize \picsize\fi
\epsfbox{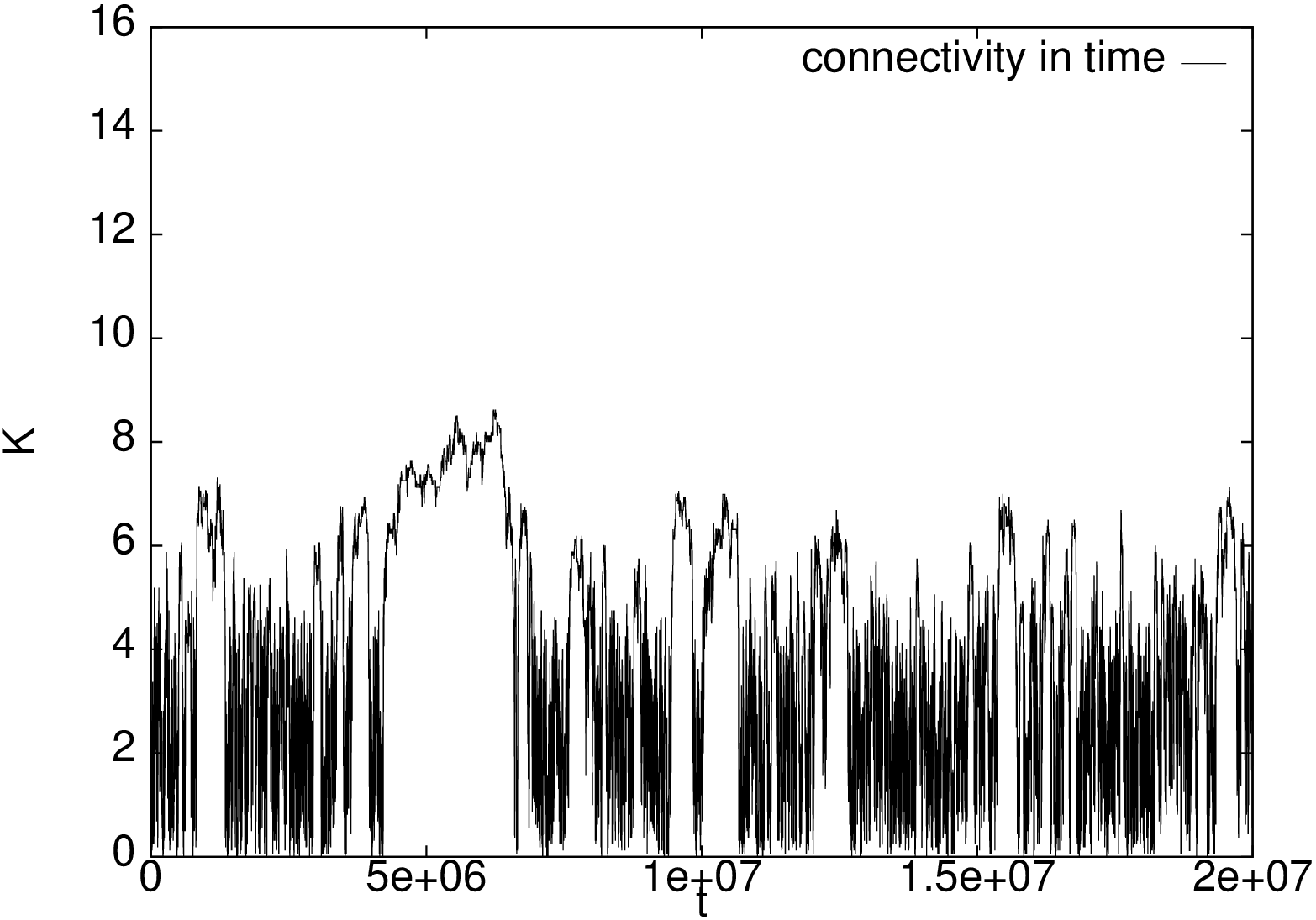}}}\fi
\end{figure}
\begin{figure}[htb]
\let\picnaturalsize=N
\def\picsize{85mm}
\def\picfilename{punctuationsfig1b.ps}
\ifx\nopictures Y\else{\ifx\epsfloaded Y\else\input epsf \fi
\let\epsfloaded=Y
\centerline{\ifx\picnaturalsize N\epsfxsize \picsize\fi
\epsfbox{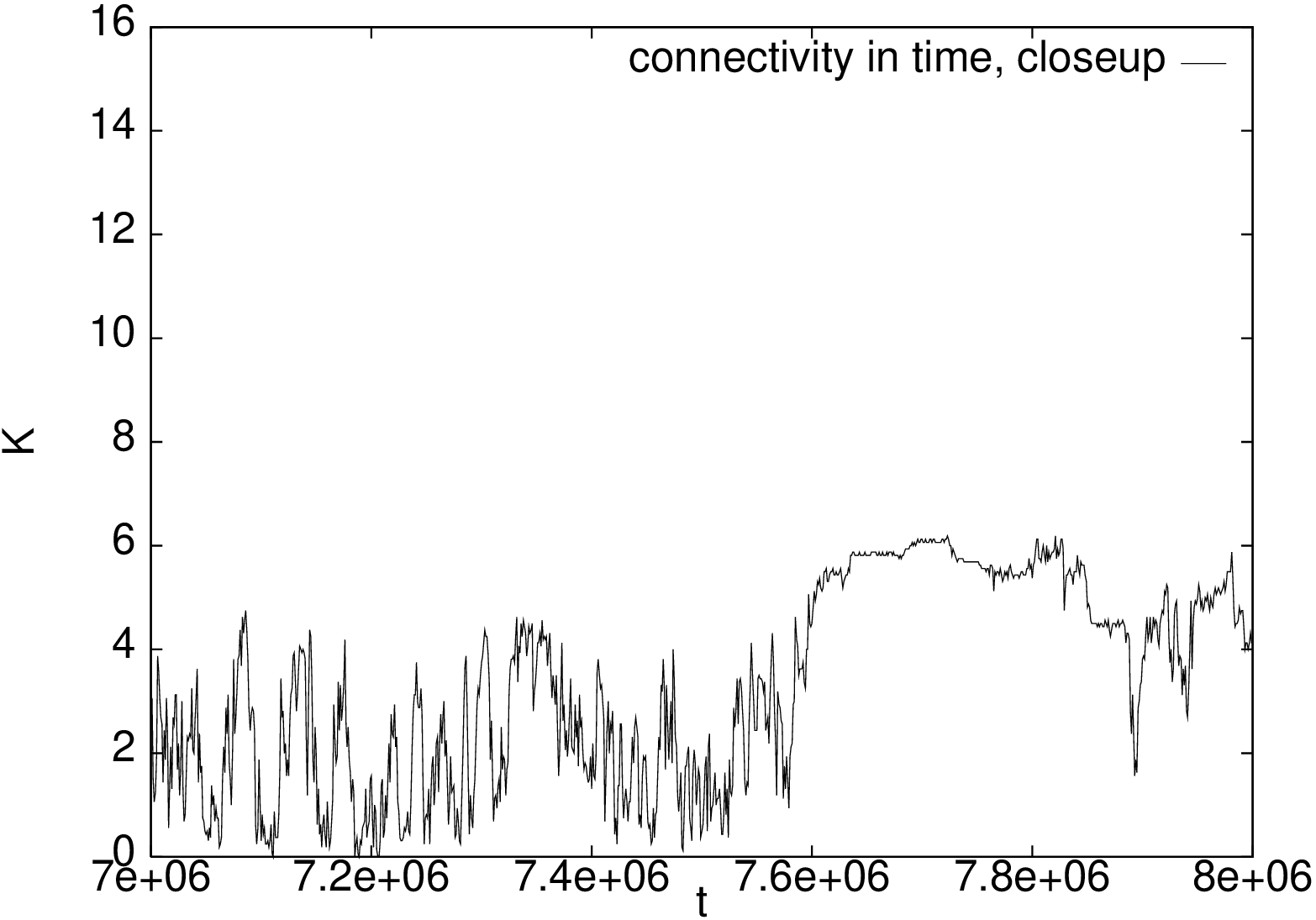}}}\fi
\end{figure}
\begin{figure}[tbh]
\caption{
Evolution of the Boolean network connectivity with time (a) and 
closeup on a part of the connectivity evolution (b). 
Note that periods of approximate stasis and sudden punctuations appear 
on both timescales.  
A single network of size  $N=16$ has been simulated starting
with an initial average connectivity of $K=1$ active inputs per node.
The connectivity matrix as well as the 
Boolean updating matrix were chosen completely randomly,
with all possible Boolean rules allowed.  
Apart from the slow adjustment of Boolean rules under step 3 in the model,
the system thus evolves in a quenched ``landscape'' of Boolean rules.
The connectivity $K$ shown is directly measured from the connectivity 
matrix of the network.
The effective connectivity $[15]$, defined as $K$ minus the
number of connected inputs that do not contribute due to specific 
Boolean updating matrix entries, 
has somewhat lower values but shows similar overall behavior.
}
\end{figure}
One observes that the typical $K$ of the network
is confined to lower values than of random networks.
This is further quantified in Fig.\ 2 where the distributions
of average connectivities are displayed in the 
statistically stationary state.
\begin{figure}[htb]
\let\picnaturalsize=N
\def\picsize{85mm}
\def\picfilename{punctuationsfig2.ps}
\ifx\nopictures Y\else{\ifx\epsfloaded Y\else\input epsf \fi
\let\epsfloaded=Y
\centerline{\ifx\picnaturalsize N\epsfxsize \picsize\fi
\epsfbox{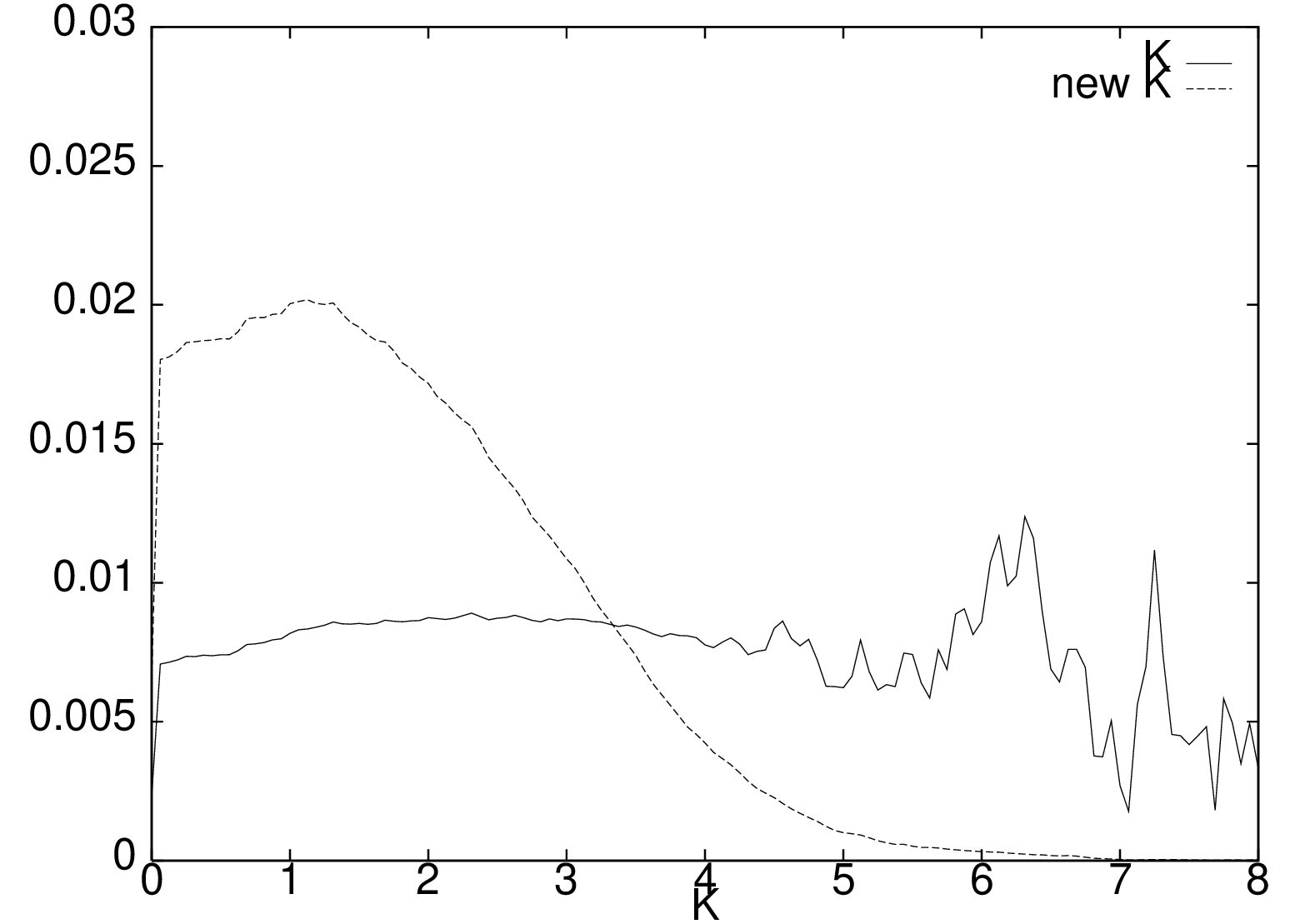}}}\fi
\end{figure}
\begin{figure}[tbh]
\caption{
Distributions of connectivity $K$ in the statistically stationary state,
obtained from the same simulation as in Figure 1.
The frequency of connectivities of the new ``species'' is shown (``new K''), 
as well as the time averaged distribution of 
connectivities of all mutated ``species''.
Note that for higher values of $K$, a mutant network is 
less likely to match the phenotype of its parent.}
\end{figure}
Notice that there are two distributions, one counting the
frequency of connectivities for all new ``species'' and one counting the
time averaged distribution.
These two distributions diverge strongly for high $K$,
because the few species with high $K$ have very long lifetimes, 
i.e., for high $K$ it is difficult to 
find mutations which do not change the
activity pattern of the networks. 
In our case, the activity pattern 
consists of the transient and the final periodic attractor following 
the given initial state. The timescale of these patterns becomes large 
for networks with high $K$, making it more difficult to 
keep the exact dynamic pattern under the mutation of a weight.
In popular terms, an increased complexity of the network makes further
evolution difficult. One may speculate that this is the reason for 
real genetic networks to keep their connectivity low: It will be
easier to evolve by increasing the number of genes $N$ at a fairly low 
connectivity level 
(the present model, however, does not consider variable $N$).   

In Fig.\ 1a we further see that marked punctuations occur,
where long periods of nearly fixed average connectivity sometimes
are interrupted by a sudden change in connectivity.
This interplay between long waiting times and short times for
actual changes is in fact observed in the fossil record.
The phenomenon has been coined ``punctuated equilibrium'' by
Gould and Eldredge \cite{GE1993}.  As also seen from Fig.\ 1b, 
the periods of stasis show a similar structure on shorter 
timescales as they do on longer timescales.
This is explored further in Fig.\ 3a where
we show this distribution averaged over the simulation.
\begin{figure}[htb]
\let\picnaturalsize=N
\def\picsize{85mm}
\def\picfilename{punctuationsfig3a.ps}
\ifx\nopictures Y\else{\ifx\epsfloaded Y\else\input epsf \fi
\let\epsfloaded=Y
\centerline{\ifx\picnaturalsize N\epsfxsize \picsize\fi
\epsfbox{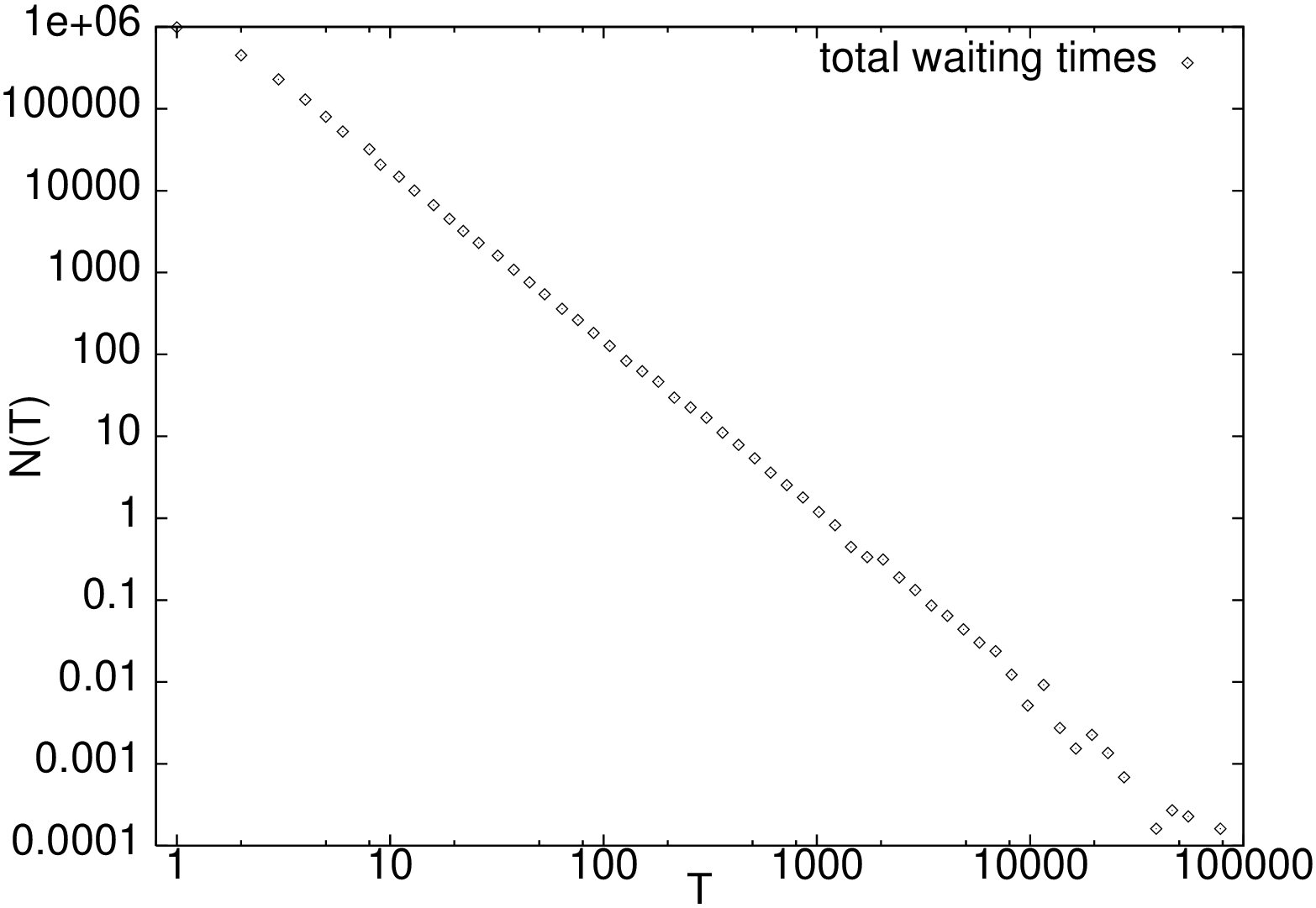}}}\fi
\end{figure}
\begin{figure}[htb]
\let\picnaturalsize=N
\def\picsize{85mm}
\def\picfilename{punctuationsfig3b.ps}
\ifx\nopictures Y\else{\ifx\epsfloaded Y\else\input epsf \fi
\let\epsfloaded=Y
\centerline{\ifx\picnaturalsize N\epsfxsize \picsize\fi
\epsfbox{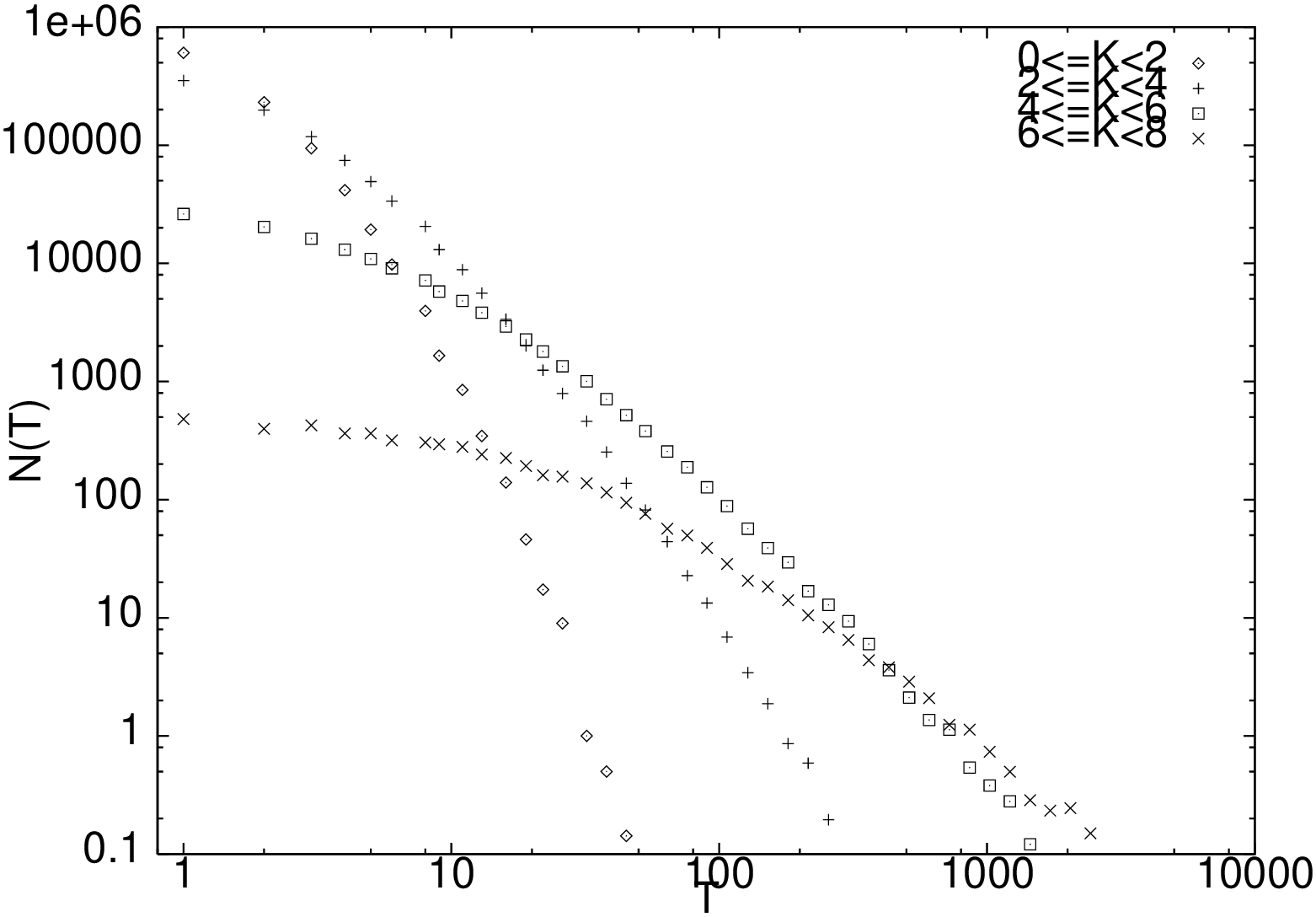}}}\fi
\end{figure}
\begin{figure}[tbh]
\caption{
Stasis time distribution in the neutral evolution of networks (a) 
and the decomposition into stasis time distributions for different 
intervals of the average connectivity (b).
The simulation is the same as in Figure 1.
While the average over all networks approximately follows a power law, 
the distributions for networks with restricted $K$ do not. 
One observes that large connectivity typically implies a lower degree of 
evolvability.
}
\end{figure}
Approximately the distribution of stasis times is $\propto 1/t^2$.
Periods of stasis at high values of $K$
can become long, which in practice calls for very
long equilibration times. 

In Fig.\ 3b we decompose the stasis time distribution into times 
obtained for different values of the average connectivity.
Again we observe that higher $K$ typically shows longer
stasis times. Remarkably, when looking at the statistics of 
a small interval of low $K$ values, we observe exponentially 
distributed stasis times. 
The power law behavior then comes about by averaging over 
the range of all $K$ values. 

In order to test for the robustness of our model 
we tried other mutation rules (again without any evolutionary pressure,  
i.e., symmetric in adding and removing weights). 
In one variant a daughter network was created by 
a) adding or b) removing a weight in the coupling matrix at random, 
with $p=1/2$ each, 
thus allowing for $K$-changing mutations only. 
In another variant a daughter network was created by independently 
adding a random weight with 
$p=1/2$
and removing a random weight with 
$p=1/2$.  
We also tested a scenario where we demanded complete match between 
attractors of mother and daughter for  
two different initial configurations. 
Also, we considered this case with demanding only 
partial overlap between mother and daughter,  
i.e., match in at least one of the two tested environments, only. 
Finally, we tested networks with weights $w_{ij} \in \{-1,0,1\}$ 
such that the signal transmitted from an inactive node can
differ from the value of a disconnected input node. 
In all cases our results were robust.

Let us briefly discuss the meaning of the 
stasis times and punctuations observed here. 
According to the definition of our model, 
we quantify the waiting time in terms of the number of times
mutant networks are exposed to new environments  
before a neutral mutation occurs that fulfills continuity.  
Thus they are not to be confused with the ``neutral evolution'' 
introduced by Kimura \cite{Kimura} which leads to 
waiting times consisting of a number of neutral mutations.
The genetic networks are formally defining a 
``species'' and the length of the waiting times  
indicates the ``genetic flexibility'' of a species.   

Associating the interconnectedness of the
networks with the genetic flexibility of real organisms
one may attempt to understand a puzzling decomposition
of lifetimes of species in the fossil record.
First it was noted by Van Valen \cite{VanValen} that each group of
closely related species has exponentially distributed lifetimes.
Second, an analysis of the overall distribution of genera
lifetimes, tabulated by Raup and Sepkoski \cite{Raup}, showed that this is 
rather distributed as $\propto 1/t^2$\cite{PNAS} for genera lifetimes
exceeding 10 million years.  
It is tempting to speculate
that groups of closely related species are associated to the
same genetic flexibility, and thus evolve, and eventually get extinct,
with a frequency given by this genetic flexibility.
This would explain the exponential distribution of Van Valens.
Averaging over all genetic flexibilities is then an average over
different characteristic lifetimes, and our 
simplified evolution scenario demonstrates 
how such an averaging can give an overall $1/t^2$ distribution.

The obtained $1/t^2$ scaling may be an
inherent part of our neutral evolution scenario
\cite{Flyvbjerg}.
In comparison, for evolution on fitness landscapes
one typically obtains a distribution $\propto 1/t$ 
corresponding to a sampling of waiting times for passing 
over barriers \cite{PNAS,Newman1996},  
although a $1/t^2$ distribution can be obtained
by supplementing a hill climbing concept with the assumption 
that extinction of a given species is determined
by evolutionary rates of older species \cite{Sibani}. 
Abandoning fitness landscapes, the
ecological network model of ref.\ \cite{Sole,Manrubia}
instead determines the fate of a species through 
an ecological connectivity matrix.
When this evolving network of species is assigned a connectivity 
that is comparable to (eco)system size, 
it shows a $1/t^2$ distribution of genera lifetimes.
In contrast to these macro-evolutionary
models, our study of Boolean networks only considers one species,
with a comparison to species extinction data that is
based on an extrapolation of the obtained evolutionary clock.

In conclusion we have studied evolution of  
Boolean networks in absence of any competition. 
This simplification allowed us to discuss 
how the requirement of evolving robust networks in itself may lead 
to an evolution which exhibits punctuated equilibrium.

\acknowledgements{S.B.\ thanks NORDITA, Copenhagen, for kind support and 
warm hospitality and the Deutsche Forschungsgemeinschaft 
for funding this work.}

\end{document}